\documentclass{jetpl}
\usepackage{graphicx}

\twocolumn

\lat

\title{Threshold Perturbations in Current-Carrying Superconducting
Bridges with a Finite Length near the Critical Temperature}

\rtitle{Threshold Perturbations}

\sodtitle{Threshold Perturbations in Current-Carrying
Superconducting Bridges with a Finite Length near the Critical
Temperature}

\author{P. M. Marychev\thanks{e-mail: observermp@yandex.ru}and D. Yu. Vodolazov}

\rauthor{MARYCHEV, VODOLAZOV}

\sodauthor{MARYCHEV, VODOLAZOV}

\address{Institute for Physics of Microstructures, Russian Academy of Sciences, Nizhny Novgorod, 603950 Russia}

\dates{January 26, 2016}{February 15, 2016}

\abstract{Near the critical temperature of a superconducting
transition, the energy of the threshold perturbation $\delta
F_{thr}$ that transfers a superconducting bridge to a resistive
state at a current below the critical current $I_c$ has been
determined. It has been shown that $\delta F_{thr}$ increases with
a decrease in the length of a bridge for short bridges with
lengths $L<\xi$ (where $\xi$ is the coherence length) and is
saturated for long bridges with $L\gg \xi$. At certain geometrical
parameters of banks and bridge, the function $\delta F_{thr}(L)$
at the current $I\to 0$ has a minimum at $L\sim 2-3 \xi$. These
results indicate that the effect of fluctuations on Josephson
junctions made in the form of short superconducting bridges is
reduced and that the effect of fluctuations on bridges with
lengths $\sim 2-3 \xi$ is enhanced.}

\PACS{74.25.F-, 74.40.-n, 74.78.Na}

\begin{document}

\maketitle

It is known that a superconducting state becomes unstable with
respect to infinitely small perturbations of the superconducting
order parameter $\Delta$ when the current $I$ flowing in a
superconductor is larger than a certain critical value, $I>I_c$.
However, switching to a resistive state can occur at a lower
current if the appearance of a finite perturbation in the system
is possible. This effect is well known from the theory of
Josephson junctions ~\cite{PRB-1974}. Such a switching in
superconducting bridges/wires with a finite length $L$ was studied
experimentally in ~\cite{Nature-2009,PRL-2011}. These
perturbations are due to thermal or quantum fluctuations. If
fluctuation- induced change in the order parameter $\Delta$ is
small, the superconducting system returns to the equilibrium state
without dissipation. However, if this change in $\Delta$ is
sufficiently large, instability will be developed in the
superconductor, leading to the appearance of a finite resistance
and dissipation. As a result, in the presence of a sufficiently
high current, the superconductor can be heated and switched to the
normal state. If the energy of threshold perturbation $\delta
F_{thr}$ is much higher than the thermal energy $k_BT$, the
probability of the appearance of such a perturbation owing to
thermal fluctuation is determined primarily by the Arrhenius
factor $exp(-\delta F_{thr}/k_BT)$.

It will be shown below that threshold perturbation $\delta
F_{thr}$ in bridges with the length $L<\xi$ increases rapidly with
a decrease in $L$ because of enhanced suppression of
superconductivity in banks. For this reason, Josephson junctions
based on short bridges/constrictions are more stable with respect
to fluctuations at a decrease in the length of the
bridge/constriction. At the same time, small $\delta F_{thr}$ is
necessary in some other situations. In particular, this is
important when studying macroscopic quantum tunneling in
superconducting systems \cite{Arutyunov}. As will be shown below,
sufficiently narrow bridges with a length of about 2-3$\xi$ have
the minimum $\delta F_{thr}$. Consequently, they are more
preferable as compared to shorter or longer bridges for their use
in devices based on quantum tunneling between different states
(e.g., so-called flux qubits \cite{Mooij}).

To calculate threshold perturbation, it is necessary to determine
a saddle-point state in the system nearest in energy to the ground
state. For a long ($L \gg \xi$, where  $\xi$ is the coherence
length) one-dimensional (transverse dimensions smaller than $\xi$)
superconducting bridge, such a problem was solved in well-known
work ~\cite{PhysRev-164}. It was found that threshold perturbation
(saddle state) corresponds to a partial suppression of the
superconducting order parameter in a finite segment of the bridge
with dimensions of about $\xi$, and the amplitude of suppression
increases with a decrease in the flowing current. Langer and
Ambegaokar ~\cite{PhysRev-164} obtained the dependence of the
energy of threshold perturbation on the applied current. It is
described well by the expression~\cite{PRB-2003}

\begin{eqnarray}
 \label{df-LA}
\delta F_{LA}=\frac{4\sqrt{2}}{3}F_0 \left
(1-\frac{I}{I_{dep}}\right)^{5/4}
\\
\nonumber
 =\frac{\sqrt{6}}{2}\frac{I_{dep}\hbar}{e}\left
(1-\frac{I}{I_{dep}}\right)^{5/4},
\end{eqnarray}
where $F_0=\Phi_0^2S/32\pi^3\lambda^2\xi$, $\Phi_0$ is the
magnetic flux quantum, $S=wd$ is the area of the cross section of
the bridge with the width $w$ and thickness $d$, $\lambda$ is the
London penetration depth of the magnetic field, and
$I_{dep}=2I_0/3\sqrt{3}$ ($I_0=c\Phi_0S/8\pi^2\lambda^2\xi$) is
the depairing current in the Ginzburg--Landau model, which
coincides with the expected critical current of the long ($L \gg
\xi$) bridge.

We calculate the energy of threshold perturbation for the
superconducting bridge with an arbitrary length $L$, which can be
both smaller and larger than $\xi$. This problem is of interest
because of the development of technologies and the appearance of
superconducting bridges with a length of about the coherence
length \cite{Nature-2009,PRL-2011,Chu}. As in~\cite{PhysRev-164},
we use the Ginzburg--Landau model; therefore, our results are
applicable only near $T_c$. We find that the current dependence of
$\delta F_{thr}$ varies smoothly from~\eqref{df-LA} for bridges
with the length $L \gg \xi$ to the expression $\delta
F_{thr}=\hbar I_c (1-I/I_c)^{3/2}/e$ for bridges with the length
$L\ll\xi$, where $I_c=I_0\xi/L$ is the critical current of the
short bridge \cite{JETP-1969}. In the latter case, the current
dependence of $F_{thr}(I)$ coincides with the known result for
Josephson junctions with a sinusoidal current-phase relation
~\cite{PRB-1974}, where $I_c$ is the critical current of the
junction. Furthermore, we found that the suppression of the
superconducting order parameter in banks of short bridges is of
great importance: it is responsible for the dependence of $\delta
F_{thr}$ on the length of the bridge and the width of banks. In
the onedimensional model, we obtained the dependence $\delta
F_{thr}(I=0) \sim 1/L$ for a short bridge with $L<\xi$. In the
two-dimensional model, we found a region of the parameters where
$\delta F_{thr}(I=0)$ depends nonmonotonically on the length of
the bridge and reaches a minimum at $L \sim 2-3 \xi$. Our results
can be used to analyze experimental data on the switching current
of short superconducting bridges/wires and fluctuation resistance
of bridges at temperatures near $T_c$.

\begin{figure}[hbt]
 \begin{center}
\includegraphics[width=0.8\linewidth]{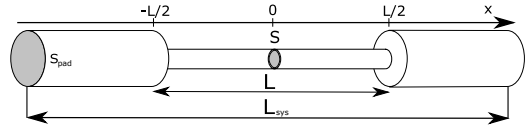}
 \caption{\label{Fig:wire}
Fig. 1. Superconducting bridge with the area of cross section $S$
and length $L$connecting superconducting banks with the area of
cross section $S_{pad}$.}
 \end{center}
\end{figure}

We consider a model system consisting of the superconducting
bridge with the area of cross section $S$ and length $L$, which
connects two superconducting banks whose cross section has the
area $S_{pad}$ (Fig.~\ref{Fig:wire}). To determine the energy of
threshold perturbation transferring the superconducting bridge to
the resistive state, we use the Ginzburg –- Landau model. To
determine $\delta F_{thr}$, it is necessary to find the state of
the system corresponding to the local minimum (extremum) of the
free energy. The saddle-point state, as well as the ground state,
can be found from the solution of the Ginzburg –- Landau equation:
\begin{equation}
 \label{GL}
   \xi^2_{GL}(0)\nabla^2\Delta+(1-T/T_c-|\Delta|^2/\Delta^2_{GL}(0))\Delta=0,
\end{equation}
where $\xi_{GL}(0)$ and $\Delta_{GL}(0)$ -- are the coherence
length and superconducting order parameter in the Ginzburg –-
Landau model at zero temperature \cite{Tinkham_book}.

For the superconducting system (see~\ref{Fig:wire}) with the
maximum characteristic transverse dimension
$d_{pad}\sim\sqrt{S_{pad}}\ll\xi$, the problem can be considered
as one-dimensional and only the dependence on the longitudinal
coordinate x is taken into account. In this case, the
dimensionless Ginzburg -- Landau equation has the form (the
solution is sought in the form $\Delta(x)/\Delta_{GL}=
f(x)exp(i\varphi(x))$)
\begin{equation}
 \label{one-dim-GL}
   \frac{d^2f}{dx^2}-\frac{j^2}{f^3}+f-f^3=0,
\end{equation}
where the condition of the constant current in the system,
$I=const$, is used (here, $j=f^2d\varphi/dx=I/S$ is the current
density in the bridge and $j=I/S_{pad}<I/S$ is the current density
in banks). In ~\eqref{one-dim-GL}, the magnitude of the
superconducting order parameter $f$, length, and current density
are measured in units of
$\Delta_{GL}=\Delta_{GL}(0)\sqrt{1-T/T_c}$ ,
$\xi=\xi_{GL}(0)/\sqrt{1-T/T_c}$, and $I_0/S$, respectively.

Equation~\eqref{one-dim-GL} should be supplemented with boundary
conditions at the ends of the bridge:
\begin{subequations}
 \label{new-bound}
\begin{equation}
  \label{pad-bridge-bound-dif}
   \frac{df^L}{dx}\bigg|_{-\frac{L}{2}}=\frac{S}{S_{pad}}\frac{df^{C}}{dx}\bigg|_{-\frac{L}{2}}=\frac{S}{S_{pad}}\frac{df^{C}}{dx}\bigg|_{\frac{L}{2}}=\frac{df^{R}}{dx}\bigg|_{\frac{L}{2}},
     \end{equation}
      \begin{equation}
      \label{pad-bridge-bound-f}
       f^L\big|_{-\frac{L}{2}}=f^C\big|_{-\frac{L}{2}}=f^C\big|_{\frac{L}{2}}=f^R\big|_{\frac{L}{2}},
       \end{equation}
       \begin{equation}
        \label{sys-bound}
        f^L\big|_{-\frac{L_{sys}}{2}}=f^R\big|_{\frac{L_{sys}}{2}}=1,
        \end{equation}
\end{subequations}
where $f^{L}, f^{C}$ and $f^{R}$ are the magnitudes of the order
parameter in the left bank, bridge, and right bank, respectively.

Condition~\eqref{pad-bridge-bound-dif} appears from the variation
of the Ginzburg –- Landau functional for the superconductor with
the cross section depending on $x$ (which is responsible for the
appearance of the derivative $d/dx(S(x)df/dx)$] in the Ginzburg –-
Landau equation). This condition is exact in the case of a
continuous change in $S$ slow at the scale $\xi$, where the
dependence of $f$ on the transverse coordinate can be neglected.
In our model, this change is stepwise. Consequently, the ratio
$S/S_{pad}$ here is not the actual ratio of the areas of the cross
sections but is a reference parameter characterizing a change in
the derivative of the function $f$ in the $x$ direction at the
transition through the bank –- bridge interface. We also assume
that the entire system is connected to wider banks (located at
$x=\pm L_{sys}/2$)), where the current density is almost zero and
the order parameter reaches its equilibrium value $f=1$. In order
to exclude the effect of these banks on the transport
characteristics of the bridge, we set $L_{sys}-L=20 \xi$ in
numerical calculations. The energy of threshold perturbation can
be determined using the expression
\begin{equation}
 \label{fluct-energy}
\frac{\delta
F_{thr}}{F_0}=F_{saddle}-F_{ground}-2\frac{I}{I_0}\delta\varphi,
\end{equation}
where $\delta\varphi$ is the additional phase difference between
the ends of the bridge appearing in the saddle-point state and
$F_{saddle}$ and $F_{ground}$ are the dimensionless free energies
of the saddle-point and ground states, respectively:
\begin{equation}
 \label{free-energy}
   F_{saddle,ground}=-\frac{1}{2}\int f^4
   dx.
\end{equation}

Equation~\eqref{one-dim-GL} with boundary
conditions~\eqref{new-bound} was solved numerically for arbitrary
$L$ values and analytically in the limit $L\ll\xi$. In the
numerical solution, we used the relaxation method: the time
derivative $\partial f/\partial t$ was added to Ginzburg -- Landau
equation~\eqref{one-dim-GL} and iterations were performed until
the time derivative became zero within a given accuracy. To find
the saddle- point state, we used the numerical method proposed in
\cite{Vodolazov_PRB}: at a given current, we fixed the magnitude
of the order parameter $f(0)$ in the center of the bridge,
allowing variations of $f$ at all other points. The state with the
minimum fixed $f(0)$ value for which a steady-state solution
exists is a saddle-point state. In the case of long bridges, this
numerical method gives $\delta F_{thr}$ values coinciding with Eq.
~\eqref{df-LA}.

To analytically find the energy of the saddle-point state, we take
into account that the order parameter varies rapidly at scales
much smaller than $\xi$. Therefore, the linear and cubic terms can
be neglected in Eq.~\eqref{one-dim-GL} for a short bridge. In this
case, we arrive at the equation
\begin{equation}
 \label{Laplace-eq}
   \frac{d^2f}{dx^2}-\frac{j^2}{f^3}=0,
\end{equation}
which has the first integral
\begin{equation}
 \label{first-int-Lap}
  \frac{1}{2} \left(\frac{df}{dx}\right)^2+\frac{j^2}{2f^2}=E,
\end{equation}
and the solution
\begin{eqnarray}
 \label{Laplace-time}
   x=\frac{1}{2}\int_{u_1}^u
   \frac{du}{\sqrt{2Eu-j^2}}=
    \\
    \nonumber
   =\frac{1}{\sqrt{2E}}\left(\sqrt{u-\frac{j^2}{2E}}-\sqrt{u(0)-\frac{j^2}{2E}}\right).
\end{eqnarray}

Here, $u(x)=f^2(x)$. Owing to the symmetry of the system,
$\frac{du}{dx}\bigg|_{x=0}=0$. At this step, we assume that
variations of $f$ are small in banks and use the boundary
condition $u\left(L/2\right)=u\left(-L/2\right)=1$ to find the
constant $E$:

\begin{eqnarray}
 \label{f(s)}
   f=\sqrt{2E_{\pm}x^2+\frac{j^2}{2E_{\pm}}},
   \\
   \label{quasienergy-1}
    E_{\pm}=\frac{1\pm\sqrt{1-\left(\frac{I}{I_{c}}\right)^2}}{L^2},
\end{eqnarray}
where $E_{+}$ and $E_{-}$ correspond to the saddle-point and
ground states, respectively, and $I_c$ is the critical current of
the short bridge~\cite{JETP-1969}.

It is fundamentally important to take into account change in
$\Delta$ in the banks when determining the energy of the
saddle-point state in the case of short bridges. Otherwise, fixing
$\Delta$ in the banks, as in the problem of the critical current
of bridges \cite{JETP-1969}, one can find (from solutions
presented below; see Eq.~\eqref{df-true}) that the energy of the
saddle-point state is negative in a wide range of the current
$I<I_c$.

\begin{figure}[hbt]
 \begin{center}
\includegraphics[width=1.0\linewidth]{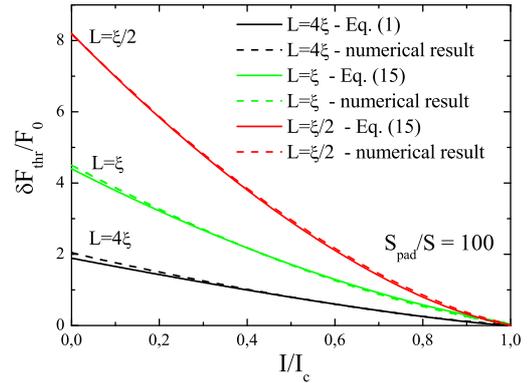}
 \caption{\label{Fig:wire_vs_jos}
Fig. 2. Current dependences of the energy of threshold
perturbation for bridges with different lengths at the ratio of
areas $S_{pad}/S=100$. The dashed lines correspond to numerical
calculations and the solid lines are obtained from Eq.
~\eqref{df-LA} for a bridge with the length $L=4\xi$ and from Eq.
~\eqref{df-true} for bridges with $L=\xi$ and $\xi/2$.}
 \end{center}
\end{figure}

We seek the solution in the banks in the form $f=1-f_1$, where
$f_1\ll 1$, and neglect the depairing effect of the current. Then,
Eq.~\eqref{one-dim-GL} for $f_1$ in the range $|x|>L/2$ becomes
\begin{equation}
 \label{one-dim-GL-f1}
  \frac{d^2f_1}{dx^2}-2f_1=0.
\end{equation}
with the solution
\begin{equation}
 \label{f1}
   f_1=Ce^{\pm\sqrt{2}(x\pm L/2)},
\end{equation}
where the signs $+$ and $-$ correspond to the left and right
banks, respectively. The constant $C$ is determined from boundary
conditions ~\eqref{new-bound}. When the current density in the
banks is $j \ll 1$ and the ratio of the cross sections is
$S/S_{pad}\ll 1$, Eq.~\eqref{f(s)} can be used for $f$ and the
constant $C$ is determined as
\begin{equation}
 \label{const}
   C=\frac{S}{S_{pad}}\sqrt{E_{\pm}-\frac{j^2}{2}}.
\end{equation}

Taking into account a decrease in the order parameter in the banks
in Eq. \eqref{free-energy}, we obtain the following expression for
the energy of threshold perturbation:
\begin{eqnarray}
 \nonumber
  \frac{\delta F_{thr}}{F_0}=2\sqrt{2}\frac{\xi}{
  L}\left(\sqrt{1+\sqrt{1-\gamma^2}-\frac{\gamma^2}{2}}-\right.
  \\
 \nonumber
  \left.-\sqrt{1-\sqrt{1-\gamma^2}-\frac{\gamma^2}{2}}\right)+\frac{2}{5}\frac{L}{\xi}\sqrt{1-\gamma^2}-
   \\
    \label{df-true}
  -4\gamma\frac{\xi}{L}\arccos(\gamma),
\end{eqnarray}
where $\gamma=I/I_c$. If a decrease in $\Delta$ in the banks is
disregarded, the first term in Eq. \eqref{df-true} is absent and
$\delta F_{thr}<0$ is in a wide range of currents.

Figure 2 shows the results of the numerical calculation $\delta
F_{thr}$ for bridges with different lengths in comparison with the
results obtained by expressions ~\eqref{df-LA}
and~\eqref{df-true}. It is seen that $\delta F_{thr}$ for a bridge
with the length $L=4\xi$ is well reproduced by Eq.~\eqref{df-LA},
whereas Eq.~\eqref{df-true} almost exactly reproduces the
numerical results for the bridge with the length $L=\xi$.
Expression~\eqref{df-true} for short bridges with lengths $L\ll
\xi$ is close to the approximation expression
\begin{equation}
 \label{df-appr}
\delta F_{thr}=\frac{4\xi}{L}F_0(1-I/I_c)^{3/2}=\frac{I_c
\hbar}{e}(1-I/I_c)^{3/2}
\end{equation}
which coincides with the known result~\cite{PRB-1974} following
from the theory of Josephson junctions with the sinusoidal
current–phase relation if $I_c$ is treated as the critical current
of a Josephson junction.
\begin{figure}[hbt]
 \begin{center}
\includegraphics[width=1.0\linewidth]{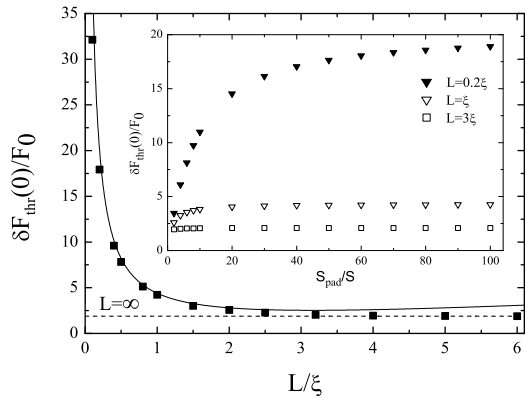}
 \caption{\label{Fig:df-zero}
Fig. 3. Energy of threshold perturbation versus the length of the
bridge in the limit $I \to 0$ at $S_{pad}/S=100$. The closed
squares are the results of the numerical calculations and the line
corresponds to \eqref{df-zero}. The inset shows $\delta
F_{thr}(0)$ versus $S_{pad}/S$ for bridges with different lengths
(onedimensional model).}
 \end{center}
\end{figure}

Expression~\eqref{df-true} was obtained under the assumption that
$C\ll 1$ and $f_1\ll 1$, which are ensured by the condition
\begin{equation}
 \label{cond-appl}
   \frac{S}{S_{pad}}\frac{\xi}{L}\ll 1.
\end{equation}
Our numerical calculations show that the current dependence of the
ratio $\delta F_{thr}(I/I_c)/\delta F_{thr}(0)$ varies slightly
and is determined primarily by the length of the bridge even when
condition\eqref{cond-appl} is invalid and $\delta F_{thr}(0)$
depends on the ratio $S_{pad}/S$ (see inset in Fig. 3).

If Eq. \eqref{df-true} is formally used for bridges with an
arbitrary length, the dependence $\delta F_{thr}(L)$ at $I\to 0$
has the form
\begin{equation}
 \label{df-zero}
  \frac{\delta F_{thr}}{F_0}(I\to0)=\frac{4\xi}{ L}+\frac{2L}{5\xi}.
\end{equation}
According to Eq. \eqref{df-zero}, $\delta F_{thr}$ should have a
minimum at $L=\sqrt{10} \xi \simeq 3 \xi$. However, numerical
calculations within the one-dimensional model do not confirm this
result (see Fig. 3). With an increase in the length of the bridge,
$\delta F_{thr}$ decreases monotonically, approaching the known
value $\delta F_{thr}(0)/F_0=4\sqrt{2}/3\simeq 1.89$ at $L \gg
\xi$ (see Eq.~\eqref{df-LA}).

\begin{figure}[hbt]
 \begin{center}
\includegraphics[width=0.8\linewidth]{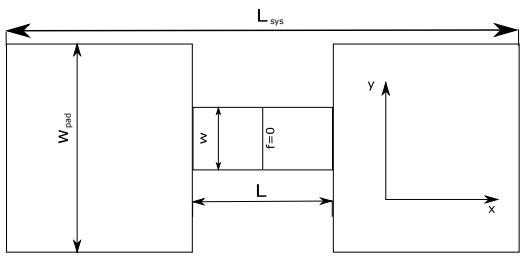}
 \caption{\label{Fig:2D}
Fig. 4. Two-dimensional superconducting bridge with the length $L$
and width $w$ with banks of the width $w_{pad}$. The thicknesses
of the bridge and banks coincide with each other; $L_{sys}-L=6$.}
 \end{center}
\end{figure}

However, the sizes of the bridge and banks at which the dependence
$\delta F_{thr}(L)$ at $I\to 0$ is nonmonotonic can be found
beyond the one-dimensional model. To this end, we considered a
two-dimensional model system shown in Fig.~\ref{Fig:2D}. This
model implies the numerical solution of the two-dimensional
Ginzburg.Landau equation
\begin{equation}
 \label{two-dim-GL-f}
  \frac{\partial^2f}{\partial x^2}+\frac{\partial^2f}{\partial y^2}+f-f^3=0,
\end{equation}
with fixed $f(x=\pm L_{sys}/2,y)=1$, normal derivative $\partial
f/\partial n=0$ at the other edges of the superconducting system,
and additional condition $f(x=0,y)=0$ (see Fig.~\ref{Fig:2D})
corresponding to the saddle-point state at $I\to 0$.

We considered various $w_{pad}$, $w$, and $L$. It is seen in Fig.
5 that the dependence $\delta F_{thr} (L)$ at a sufficiently small
width of the bridge ($w=\xi/5$ for the parameters under
consideration) has a minimum at $L\simeq 2-3 \xi$.

\begin{figure}[hbt]
 \begin{center}
\includegraphics[width=1.0\linewidth]{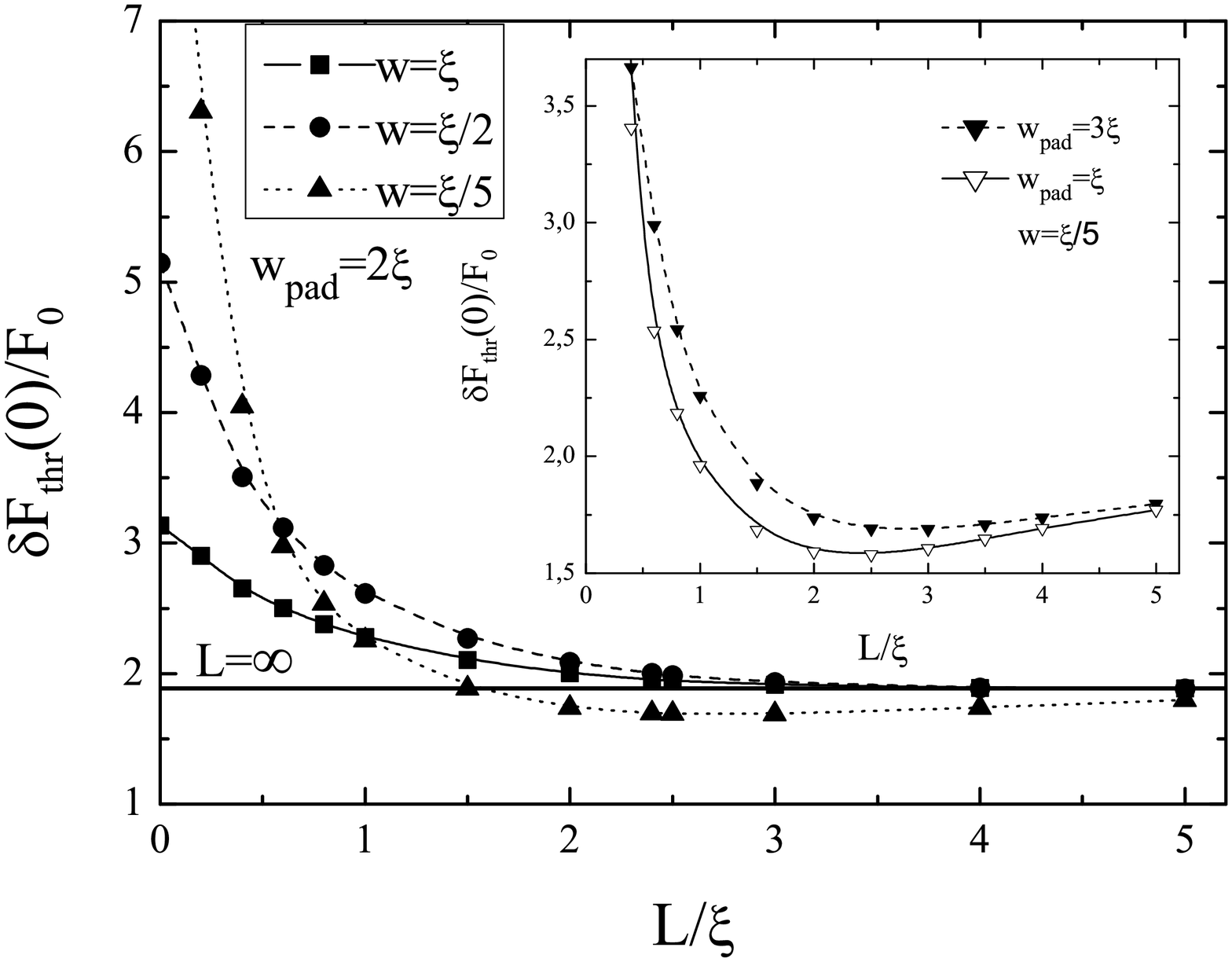}
 \caption{
Fig. 5. Energy of threshold perturbation versus the length of the
bridge in the limit $I\to 0$ at different widths of the bridge and
banks calculated within the two-dimensional model. The thicknesses
of the bridge and banks coincide with each other.}
 \end{center}
\end{figure}

We also analyzed the dependence of $\delta F_{thr}(0)$ on the
width of the banks in the two-dimensional model (the data are
shown in Fig. 6). As in the one-dimensional model, the energy of
threshold perturbation becomes independent of the ratio
$w_{pad}/w$ when the width of the bridge becomes much smaller than
$w_{pad}$. However, for the chosen geometry (see Fig. 4),
saturation occurs at smaller $w_{pad}/w$ values (cf. inset in Fig.
3).
\begin{figure}[hbt]
 \begin{center}
\includegraphics[width=1.0\linewidth]{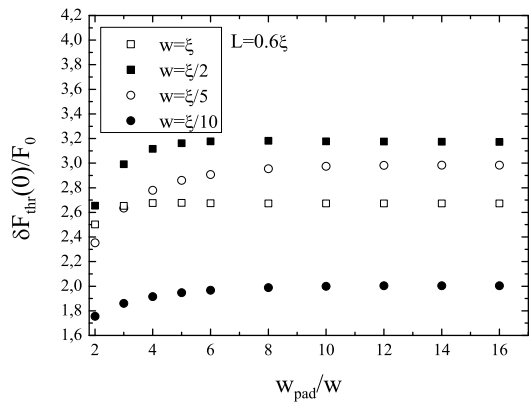}
 \caption{
Fig. 6. Energy of threshold perturbation at zero current
calculated for the bridge with the length $L=0.6 \xi$ within the
two-dimensional model for different widths of the bridge and
banks.}
 \end{center}
\end{figure}

Using the reasons of Little \cite{Chu,Little}, we can estimate a
finite (owing to thermal fluctuations) resistance of the short
bridge $R$ at low currents by the expression
\begin{eqnarray}
 \label{resistance}
R=R_n exp\left(-\frac{\delta F_{thr}(0)}{k_BT}\right)=R_n exp
\left(-\frac{I_0 \hbar}{ek_B T} \frac{\xi}{L}\right)
\\
\nonumber =R_n exp \left(-\frac{R_q}{R_n}\frac{2.66\Delta(0)}{k_B
T}\left(1-\frac{T}{T_c}\right)\right),
\end{eqnarray}
where $R_n$ is the resistance of the bridge in the normal state,
$R_q=\pi\hbar/2e^2$ is the resistance quantum and $\delta
F_{thr}(0)$ is given by Eq.~\eqref{df-appr}. According to
Eq.~\eqref{resistance}, $R$ decreases exponentially with a
decrease in the length of the bridge. This effect is due to the
suppression of $\Delta$ in the banks in the saddle-point state,
which is stronger for a shorter bridge. Thus, the banks are
responsible for a decrease in $R$, but only for a sufficiently
short bridge with $L<\xi$. In the case of a long bridge, the banks
either do not affect the resistance of the bridge or increase it.
The latter effect is possible only for sufficiently narrow bridges
(see Fig. 5), for which a decrease in $\delta F_{thr}$ owing to a
decrease in the length of the bridge is not compensated by an
increase in $\delta F_{thr}$, which is due to the suppression of
$\Delta$ in banks.

It is worth noting that the fluctuation resistance of the bridge
depends strongly not only on its length but also on the size of
the banks in view of the dependence of $\delta F_{thr}(0)$ on
$S_{pad}/S$ (see inset in Fig. 3 and Fig. 6). The critical current
depends also on the ratio $S_{pad}/S$. In particular, within the
one-dimensional model, it is easy to show that
\begin{equation}
I_c=I_0\frac{\xi}{L}\left(1-\sqrt{2} \frac{S}{S_{pad}}
\frac{\xi}{L}\right)
\end{equation}
under condition~\eqref{cond-appl}. However, since $\delta
F_{thr}(0)$ appears in Eq.~\eqref{resistance} in the exponential,
variations of the size of the banks affect the fluctuation
resistance $R$ more strongly than its critical current.

This work was supported by the Russian Foundation for Basic
Research (project no. 15-42-02365).

\end{document}